\documentclass[sigconf, twocolumn]{acmart}

\usepackage{xcolor}
\usepackage{amsmath}         
\usepackage{graphicx}        
\usepackage{booktabs}       
\usepackage{geometry}        
\usepackage{enumitem}

\title{Anchoring Trends: Mitigating Social Media Popularity Prediction Drift via Feature Clustering and Expansion}

\author{Chia-Ming Lee}
\affiliation{
  \institution{Institute of Data Science\\National Cheng Kung University}
  \city{Tainan}
  \country{Taiwan}
}
\author{Bo-Cheng Qiu}
\affiliation{
  \institution{Institute of Data Science\\National Cheng Kung University}
  \city{Tainan}
  \country{Taiwan}
}
\author{Cheng-Jun Kang}
\affiliation{
  \institution{Institute of Data Science\\National Cheng Kung University}
  \city{Tainan}
  \country{Taiwan}
}
\author{Yi-Hsuan Wu}
\affiliation{
  \institution{Institute of Data Science\\National Cheng Kung University}
  \city{Tainan}
  \country{Taiwan}
}
\author{Jun-Lin Chen}
\affiliation{
  \institution{Department of Statistic\\National Cheng Kung University}
  \city{Tainan}
  \country{Taiwan}
}
\author{Yu-Fan Lin}
\affiliation{
  \institution{Miin Wu School of Computing\\National Cheng Kung University}
  \city{Tainan}
  \country{Taiwan}
}
\author{Yi-Shiuan Chou}
\affiliation{
  \institution{Department of Computer Science and Information Engineering, National Cheng Kung University}
  \city{Tainan}
  \country{Taiwan}
}
\author{Chih-Chung Hsu}
\affiliation{
  \institution{Institute of Intelligent Systems\\College of AI, National Yang Ming Chiao Tung University}
  \city{Tainan}
  \country{Taiwan}
}

\begin{abstract}

Predicting online video popularity faces a critical challenge: \textbf{prediction drift}, where models trained on historical data rapidly degrade due to evolving viral trends and user behaviors. To address this temporal distribution shift, we propose an \textbf{Anchored Multi-modal Clustering and Feature Generation (AMCFG)} framework that discovers temporally-invariant patterns across data distributions. Our approach employs multi-modal clustering to reveal content structure, then leverages Large Language Models (LLMs) to generate semantic \textbf{Anchor Features}—high-level concepts such as audience demographics, content themes, and engagement patterns—that transcend superficial trend variations. These semantic anchors, combined with cluster-derived statistical features, enable prediction based on stable principles rather than ephemeral signals. Experiments demonstrate that AMCFG significantly enhances both predictive accuracy and temporal robustness, achieving superior performance on out-of-distribution data and providing a viable solution for real-world video popularity prediction.

\end{abstract}

\ccsdesc[300]{Computing methodologies~Social Media Computing}

\keywords{Social Media Popularity Prediction, Multi-modal learning, Test-time Adaptation, Feature Clustering, Large Language Models}

\begin{document}
\maketitle

\begin{figure}
		\centering
		\includegraphics[width=0.5\textwidth]{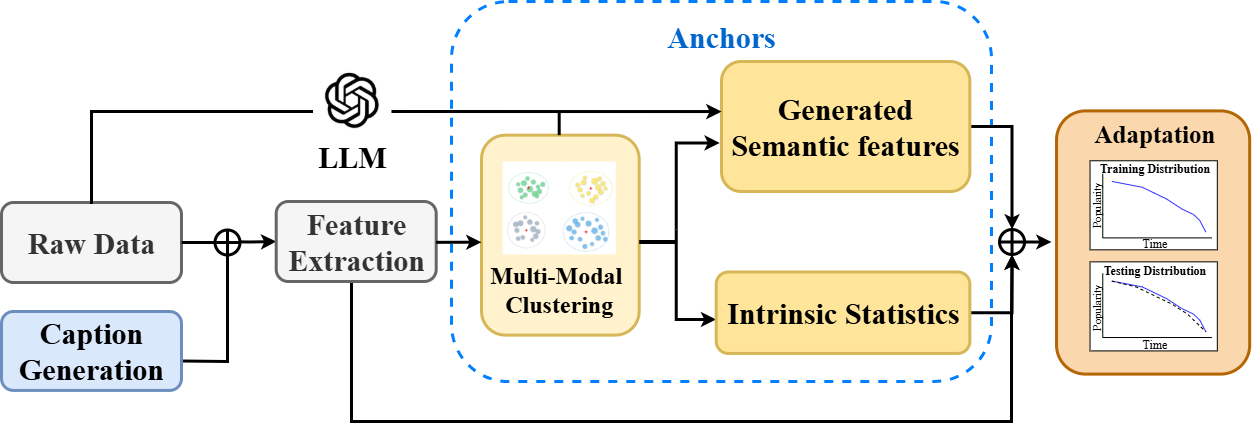}
		\caption{\small {\textbf{Conceptual illustration of anchored clustering approach for enhancing stability in social media popularity prediction.}}
  }

		\label{fig:flowchart}	
\end{figure}

\section{Introduction}

With the explosive growth of user-generated content on social media platforms like TikTok, Instagram, and YouTube, understanding the dynamics of the popularity of content has become critical for creators, advertisers, and platforms alike. The prediction of social media popularity (SMP), which predicts which posts will go viral, has immense commercial and academic value, driving research toward robust and accurate prediction models.

Initial SMP research focused on static images using tree-based models~\cite{2022Catboost,hsu2017social,hsu2018social,hsu2019social,hsu2020social}, later evolving toward Transformer-based architectures and Vision-Language Models~\cite{radford2021learning,2022Chen,chen2023double,2023sota,2022clip_visualtext,prompthsi,hsu2024} that excel at multi-modal fusion. However, as content consumption shifted decisively toward dynamic videos, these image-centric approaches became insufficient for capturing temporal dynamics and richer multi-modal signals.

The Challenge of Temporal Distribution Shift. Beyond technical complexity, SMP faces a fundamental challenge: Temporal Distribution Shift. Viral trends, memes, and content styles evolve rapidly, causing models trained on historical data to learn superficial correlations with transient features—specific meme formats, trending songs, or editing styles—that quickly become obsolete. This temporal drift severely degrades model performance on future content, creating an analogous challenge to spatial domain shifts addressed in computer vision adaptation literature.

Inspiration from Test-Time Adaptation. Recent advances in Test Time Adaptation (TTA) have demonstrated powerful solutions for distribution shift through clustering-based feature anchoring \cite{tent,shot,adacontrast,csfda}. Methods like NRC \cite{NRC} leverage neighborhood consistency to discover stable structural patterns, while approaches such as pSTarC \cite{pstarc} generate invariant representations that transcend domain-specific variations. These insights from TTA suggest that robust prediction requires discovering invariant structural patterns that transcend superficial variations - a principle directly applicable to temporal drift in social media content.

From Spatial to Temporal Anchoring. Inspired by TTA's success and recent work showing that generative AI can capture underlying content essence~\cite{simtube}, we propose an Anchored Multi-modal Clustering and Feature Generation (AMCFG) framework. Our approach adapts TTA's clustering-based anchoring philosophy to temporal settings by: (1) clustering multi-modal content to reveal stable behavioral patterns, (2) extracting statistical anchors from cluster-level engagement patterns, and (3) leveraging LLMs to generate semantic anchors—high-level concepts like content themes and audience characteristics—that remain temporally invariant even as surface trends evolve. The main contributions of this study can be summarized into three points:

\raggedbottom
\vspace{0.25cm}
\begin{itemize}
    
    \item TTA-Inspired Framework: We adapt clustering-based anchoring strategies from spatial domain adaptation to address temporal drift in the prediction of popularity of social media. By drawing from TTA's success in discovering invariant patterns across domains, we establish the first framework to systematically apply clustering-based temporal anchoring to dynamic social media environments.
    
    \item Dual-Anchoring Mechanism: We introduce complementary statistical and semantic anchors that provide both numerical and conceptual stability against temporal distribution shifts. This novel approach combines cluster-derived behavioral patterns with LLM-generated semantic concepts, creating multi-level robustness against both quantitative and qualitative temporal variations.
    
    \item Empirical Validation: We demonstrate significant improvements in model robustness and accuracy on out-of-distribution data, validating the effectiveness of temporal anchoring for real-world deployment. Our comprehensive experiments on diverse datasets and participation in the SMP2025 challenge confirm practical utility for large-scale social media applications.
\end{itemize}

The remainder of this paper proceeds as follows: Section 2 reviews related work across SMP prediction, TTA methods, and LLM-based feature engineering. Section 3 establishes our theoretical foundation connecting TTA principles to temporal drift. Section 4 details the AMCFG methodology and dual-anchoring strategy. Section 5 presents experimental validation and analysis. Section 6 concludes with key insights.

\section{Related Work}

\subsection{Social Media Popularity Prediction}

Early SMP research focused primarily on static image content, establishing foundational approaches using tree-based models~\cite{hsu2017social,hsu2018social,hsu2019social,hsu2020social} and feature engineering strategies. The introduction of large-scale datasets by Wu et al.~\cite{smp1,smp2,smp3} catalyzed significant advancements in the field. Subsequent works evolved toward more sophisticated Transformer-based architectures~\cite{2022Chen,chen2023double} and Vision-Language Models that excel at multi-modal fusion~\cite{radford2021learning,2022clip_visualtext,densesr}.

Recent advances include DFT-MOVLT~\cite{chen2023double}, which employs multi-objective pre-training for enhanced performance, and TTC-VLT~\cite{2022Chen}, which combines vision and language transformers with contrastive learning. Zhang et al.~\cite{2023sota} introduced dependency-aware frameworks that model post relationships, while other works explored feature analysis and data-splitting strategies~\cite{hsu2022,hsu2024}. Hsu et al. \cite{Hsu2023,drct} and  Lai et al. \cite{Hyfea} conducted comprehensive experiments to explore user metadata. However, most existing approaches focus on static image content and fail to address the critical challenge of temporal distribution shift in dynamic video environments.

\subsection{Test-Time Adaptation and Domain Shift}

Test-Time Adaptation has emerged as a powerful paradigm for addressing distribution shifts without access to source data during adaptation. TENT~\cite{tent} pioneered entropy minimization for test-time adaptation, while SHOT~\cite{shot} introduced mutual information maximization with pseudo-labeling strategies. More recent approaches like pSTarC~\cite{pstarc} generate pseudo-source samples to guide target clustering, and NRC~\cite{NRC} leverage reciprocal neighborhood structures to discover stable feature relationships.

These TTA methods share a common insight: robust adaptation requires discovering invariant structural patterns that transcend domain-specific variations. AdaContrast~\cite{adacontrast} demonstrates the effectiveness of contrastive learning with feature banks, while methods like C-SFDA~\cite{csfda} employ curriculum learning frameworks. Our work draws inspiration from these clustering-based anchoring strategies, adapting their spatial domain shift solutions to address temporal distribution shifts in social media content.

\subsection{Large Language Models for Feature Engineering}

Traditional automated feature engineering methods rely on predefined search spaces and heuristic algorithms~\cite{autofea,openfe}, requiring substantial computational resources and often missing valuable experimental insights. Recent advances leverage LLMs' reasoning capabilities for feature generation: CAAFE~\cite{caffe} generates semantically meaningful features based on task descriptions, but is highly dependent on the context of the language and adopts greedy evaluation strategies. OCTree~\cite{octree} uses LLMs as black-box optimizers with decision tree feedback, demonstrating effective feature space navigation without manual search constraints.

SimTube~\cite{simtube} shows that LLMs can capture underlying content essence and simulate audience feedback, supporting the hypothesis that LLMs can extract high-level semantic patterns from complex data that transcend surface-level variations.

Connection to Our Work. Unlike existing LLM-based feature engineering methods that optimize immediate performance metrics, our approach leverages LLMs to discover temporal invariances from content clusters. We use LLMs not for direct feature optimization, but for extracting semantic anchors that complement statistical anchors, creating a dual-anchoring strategy inspired by TTA's multilevel robustness approaches. This focuses on temporal stability rather than current validation performance, generating features that remain robust across evolving content trends.
\begin{figure*}
		\centering
		\includegraphics[width=1\textwidth]{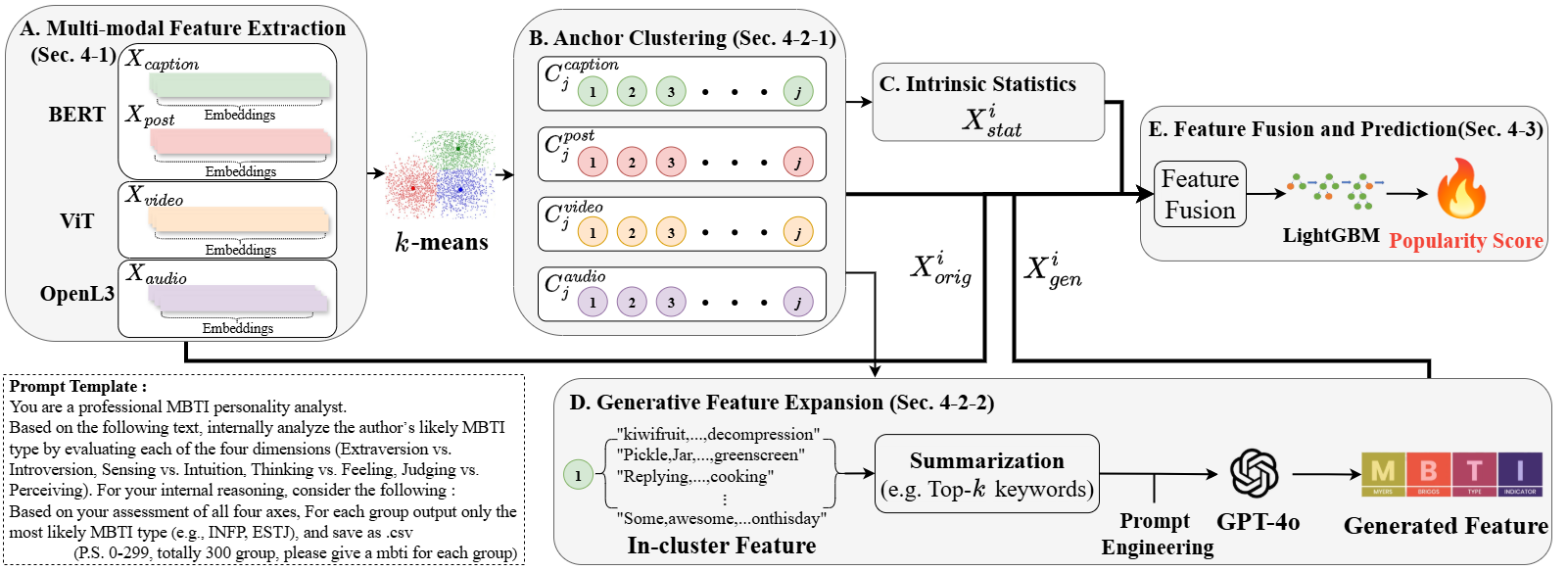}
		\caption{Overall architecture of the proposed Anchored Multi-modal Clustering and Feature Generation (AMCFG) framework. Our TTA-inspired approach leverages multi-modal clustering to discover stable behavioral patterns, then generates dual anchoring features: statistical anchors from cluster-level engagement patterns and semantic anchors from LLM-based thematic analysis. This dual-anchoring mechanism enables robust popularity prediction that transcends temporal distribution shifts in social media content.}
		\label{fig:overview}
\end{figure*}

\section{PRELIMINARY}

Inspiration from Test-Time Adaptation. Recent advances in TTA have demonstrated the power of cluster-based approaches to address challenges in the distribution shift. Methods like pSTarC~\cite{pstarc} generate pseudo-source samples to guide target clustering, while approaches such as NRC~\cite{NRC} leverage reciprocal neighborhood structures to identify stable feature representations. The high-level insight underlying these TTA methods is profound: when faced with evolving data distributions, robust prediction requires discovering invariant structural patterns that transcend superficial variations.

From Domain Shift to Temporal Drift. The core challenge in TTA---adapting models to new domains without access to source data---shares fundamental similarities with our SMP prediction task. Although TTA addresses spatial domain shifts (e.g., synthetic$\rightarrow$real images), SMP confronts temporal distribution shifts where viral trends, content styles, and user behaviors evolve rapidly over time. Both scenarios require models that can anchor predictions on stable, underlying patterns rather than transient surface features.

Clustering as a Bridge to Stability. Inspired by TTA's success in using clustering to reveal data structure, we hypothesize that content clustering can serve as a temporal anchor for SMP prediction. Just as TTA methods cluster target features to find reliable neighborhood relationships, we propose clustering multi-modal content to identify semantically coherent groups that exhibit stable popularity patterns across time periods.

Problem Formulation. In video-based SMP prediction task, a predictive model $f$ aims to predict the popularity score $\hat{y}$ from diverse post-information. This information encompasses both unstructured content including textual elements (titles, captions), visual components (video frames), and audio tracks, as well as structured metadata providing user profiles and post attributes.

However, models relying solely on these raw features $X_{\text{orig}}$ suffer from severe prediction drift---the temporal analog of domain shift in TTA. The features derived directly from the content become obsolete as trends evolve, leading to a significant increase in the prediction error $\text{Error}(y,\hat{y})$ over time. Drawing from TTA's anchoring philosophy, we propose generating two additional feature types:
\begin{enumerate}
    \item \textbf{Statistical Anchors} ($X_{\text{stat}}$): Stable behavioral patterns from cluster-level statistics
    \item \textbf{Semantic Anchors} ($X_{\text{gen}}$): High-level concepts inferred via LLM reasoning over cluster semantics
\end{enumerate}

Our framework constructs $X_{\text{final}} = \text{concat}(X_{\text{orig}}, X_{\text{stat}}, X_{\text{gen}})$ such that prediction $\tilde{y} = f(X_{\text{final}})$ satisfies:
\begin{equation}
\text{Error}(y,\tilde{y}) \ll \text{Error}(y,\hat{y}).
\end{equation}

This approach transforms the SMP prediction drift problem into a clustering-anchored stability problem, leveraging the same structural insights that make TTA methods robust across domains.

\section{Methodology}
Framework Overview. Drawing inspiration from TTA's clustering-based anchoring strategies, the AMCFG framework (Figure \ref{fig:overview}) systematically generates temporally robust features through three coordinated stages: (1) comprehensive multi-modal feature extraction from raw content, (2) modality-specific clustering to discover stable behavioral cohorts and statistical patterns, and (3) LLM-driven semantic expansion to capture high-level conceptual invariances. This approach mirrors TTA's philosophy of discovering structural stability within evolving data distributions.

\subsection{Initial Multi-modal Feature Extraction}
For each social media post $X_i$, we encode unstructured content across text, visual, and audio modalities into high-dimensional representations that serve as the foundation for subsequent clustering and anchoring.

Textual Embeddings. Textual content carries core thematic and sentiment information through both original post metadata (titles, descriptions) and VideoLLaMA 3-generated post-wise semantic features. As detailed in Table~\ref{tab:llm_prompts}, we employ VideoLLaMA3 \cite{videollama} to generate post-wise category classifications, audience demographics, and semantic descriptions for each video. These VideoLLaMA3-generated captions and categorical features, combined with original textual content, are processed using BERT-large-uncased \cite{bert}:
\begin{equation} \label{eq:text_emb}
    e_i^{\text{text}} = \text{BERT}(\text{tokenizer}(\text{caption}_i \oplus \text{description}_i \oplus \text{post content}_i)),
\end{equation}
where the \texttt{[CLS]} token representation provides $e_i^{\text{text}} \in \mathbb{R}^{d_{\text{bert}}}$, and $\text{videollama3\_features}_i$ contains the post-wise generated category classifications and semantic captions. Here, $\oplus$ denotes the concatenation of strings.

Visual Embeddings. Video content is processed using a pre-trained Vision Transformer (ViT) \cite{vit} with frame-level extraction and temporal pooling:
\begin{equation} \label{eq:vis_emb}
    e_i^{\text{vis}} = \frac{1}{T} \sum_{t=1}^{T} \text{ViT}(\text{frame}_{i,t}),
\end{equation}
where $T=256$ frames are sampled to produce $e_i^{\text{vis}} \in \mathbb{R}^{d_{\text{vit}}}$.

Audio Embeddings. Audio tracks contribute essential acoustic and musical context. We utilize Open-L3~\cite{openl3} to extract 128-dimensional features:
\begin{equation} \label{eq:audio_emb}
    e_i^{\text{audio}} = \text{Open-L3}(\text{audio}_i).
\end{equation}

From Instance Features to Structural Patterns. While these multi-modal embeddings $\{e_i^{\text{text}}, e_i^{\text{vis}}, e_i^{\text{audio}}\}$ capture rich content semantics enhanced by post-wise LLM-generated metadata, they remain vulnerable to temporal drift as individual trends evolve. Following TTA's insight that clustering reveals stable structural relationships, we next group similar content into coherent clusters to extract temporally invariant statistical and semantic anchors.

\setlist{nosep} 

\begin{table}[h]
\centering
\resizebox{0.45\textwidth}{!}{%
\begin{tabular}{|l|l|}
\hline
\textbf{Task} & \textbf{Output} \\
\hline
Video Description & Short semantic description of video content \\
\hline
Category Classification & Video category with explanation \\
\hline
Audience Analysis & Target demographics and interests \\
\hline
\end{tabular}%
}
\caption{Post-wise semantic features generated via VideoLLaMA3 prompts.}
\label{tab:llm_prompts}
\end{table}

\setlist{nosep} 
\vspace{-1.25cm}
\subsection{Dual-Anchoring Strategy: Statistical and Semantic Feature Generation}

TTA-Inspired Anchoring Framework. Inspired by TTA methods that establish stability through clustering-based feature anchoring, we propose a dual-anchoring strategy that addresses temporal drift at both numerical and conceptual levels. Just as TTA approaches like NRC \cite{NRC} leverage neighborhood consistency and pSTarC \cite{pstarc} generate invariant pseudo-features, our framework creates two complementary anchor types: \textbf{statistical anchors} that capture quantitative behavioral patterns, and \textbf{semantic anchors} that encode qualitative thematic invariances.

\subsubsection{Statistical Anchoring via Clustering}

Modality-Specific Clustering. We apply $k$-means clustering with $k=300$ clusters to each feature modality after preprocessing: SVD dimensionality reduction for textual embeddings $e_i^{\text{text}}$, direct clustering for visual/audio embeddings, and standardization for structured metadata. This generates cluster assignments:
\begin{equation} \label{eq:cluster_labels}
    \mathcal{L}_i = \{L_i^{\text{text}}, L_i^{\text{video}}, L_i^{\text{audio}}, L_i^{\text{user}}, L_i^{\text{post}}\},
\end{equation}
where $L_i^m \in \{1, 2, ..., 300\}$ denotes the cluster assignment for sample $i$ in modality $m$.

Behavioral Pattern Extraction. For each cluster $C_j^m$ (where $j \in \{1, 2, ..., 300\}$ is the cluster index and $m$ denotes modality), we compute intrinsic statistics that quantify stable engagement patterns: mean popularity $\mu_{\text{pop}}(C_j^m)$, popularity variance $\sigma_{\text{pop}}^2(C_j^m)$, and sample count $N_{\text{count}}(C_j^m)$. These metrics serve as numerical anchors analogous to TTA's neighborhood consistency measures:
\begin{equation} \label{eq:stat_features} 
\begin{aligned}     
X_{\text{stat}}^{(i)} = \text{concat}\bigg(&\{\mu_{\text{pop}}(C_{L_i^m}^m), \sigma_{\text{pop}}^2(C_{L_i^m}^m), N_{\text{count}}(C_{L_i^m}^m), L_i^m : \\     
&\quad m \in \{\text{text}, \text{video}, \text{audio}, \text{user}, \text{post}\}\}\bigg),
\end{aligned} 
\end{equation}
where $C_{L_i^m}^m$ represents the cluster in modality $m$ that sample $i$ is assigned to.
\subsubsection{Semantic Anchoring via LLM Reasoning}
\textbf{Conceptual Invariance Discovery.} While statistical anchoring captures "how clusters typically perform," semantic anchoring addresses "what makes clusters conceptually coherent." We use LLMs to analyze cluster content and generate thematic descriptions that transcend surface-level variations. For each cluster $C_j^m$ in text, visual, and audio modalities, we employ two complementary semantic analysis approaches: thematic description generation (e.g., "What is the primary theme of this cluster?") and personality profiling using GPT-4o's analysis framework. Additionally, we generate concise summarizations of cluster characteristics to capture essential semantic patterns. These analyses are encoded as:
\begin{equation} \label{eq:semantic_generation}
    d_j^m = \text{LLM}(\text{prompt}(C_j^m)), \quad 
    s_j^m = \text{SentenceTransformer}(d_j^m).
\end{equation}

Semantic Feature Integration. Post-level semantic features are constructed by retrieving cluster embeddings that incorporate both thematic and personality-based semantic characterizations:
\begin{equation} \label{eq:gen_features}
    X_{\text{gen}}^{(i)} = \text{concat}\left(\{s_{L_i^m}^m : m \in \{\text{text}, \text{video}, \text{audio}\}\}\right).
\end{equation}

Complementary Stability Mechanism. This dual-anchoring system mirrors TTA's multi-level robustness: statistical anchors provide numerical stability through historical performance patterns, while semantic anchors offer conceptual stability through abstracted themes and personality profiles. Together, they enable temporal generalization, where high-level content categories remain invariant despite evolving surface trends.

\subsection{Feature Fusion and Prediction}

Comprehensive Feature Fusion. Following TTA's principle of combining multiple stability mechanisms, we integrate our dual-anchoring features with original multi-modal representations to create a comprehensive predictive foundation:
\begin{equation} \label{eq:final_features}
    X_{\text{final}}^{(i)} = \text{concat}(X_{\text{orig}}^{(i)}, X_{\text{stat}}^{(i)}, X_{\text{gen}}^{(i)}),
\end{equation}
where $X_{\text{orig}}^{(i)}$ contains raw multi-modal features (textual, visual, audio embeddings, and metadata), $X_{\text{stat}}^{(i)}$ provides statistical anchors, and $X_{\text{gen}}^{(i)}$ contributes semantic anchors.

Robust Prediction Framework. This integration strategy mirrors TTA's approach of leveraging both local patterns (original features) and global structures (anchored features) for enhanced generalization. The resulting feature vector $X_{\text{final}}^{(i)}$ enables the model to make predictions based on immediate content characteristics while being anchored by temporally stable patterns discovered through clustering, creating a robust foundation for popularity prediction that resists temporal distribution shifts.

\section{Experiment Results and Analysis}

\subsection{Experimental Setup}

Dataset and Evaluation. We evaluate on the SMP-Video dataset containing 6K TikTok short-form videos from 4.5K users. Following TTA evaluation protocols that emphasize out-of-distribution robustness, we employ 5-fold Group K-Fold cross-validation using user identifiers to prevent data leakage, ensuring temporal generalization assessment. We use standard regression metrics: Mean Absolute Error (MAE), Mean Absolute Percentage Error (MAPE), and $R^2$ Score.
\begin{equation}
    \text{MAE}(y,\hat{y}) = \frac{1}{N} \sum_{i=1}^{N} |y_i - \hat{y}_i|;\quad 
    \text{MAPE}(y,\hat{y}) = \frac{1}{N} \sum_{i=1}^{N} \left| \frac{y_i - \hat{y}_i}{y_i} \right| \times 100\%.
\end{equation}
\begin{equation}
    R^2(y,\hat{y})= 1 - \frac{\sum_{i=1}^{N} (y_i - \hat{y}_i)^2}{\sum_{i=1}^{N} (y_i - \hat{y})^2}.
\end{equation}

Implementation Details. We employ LightGBM \cite{lightgbm}as our base predictor with \texttt{regression\_l1} objective, learning rate 0.05, and 31 leaves. For clustering, we use $k$-means with $k=300$ clusters across all modalities, applying SVD dimensionality reduction for textual embeddings and standardization for metadata features.

\begin{table}[h]
\centering
\resizebox{0.43\textwidth}{!}{%
\begin{tabular}{|c|c|c|c|}
\hline
\textbf{Comparison of $k$} & \textbf{MAPE (↓)} & \textbf{MAE (↓)} & \textbf{R²(↑)} \\
\hline
$k=100$ & 23.33\% & 1.5798 & 0.4938 \\
$k=200$ & 22.94\% & 1.5487 & 0.5161 \\
\textbf{$k=300$} & \textbf{22.53\%} & \textbf{1.4677} & \textbf{0.5535} \\
$k=400$ & 22.78\% & 1.5179 & 0.5275 \\
$k=500$ & 23.90\% & 1.5877 & 0.4835 \\
\hline
\end{tabular}%
}
\caption{Performance comparison of different setup of $k$.}
\label{tab:kmeans_effectiveness}
\vspace{-0.5cm}
\end{table}

\vspace{-0.5cm}
\subsection{Ablation Studies}

Optimal Clustering Granularity. We systematically evaluated cluster counts $k \in \{100, 200, 300, 400, 500\}$ to determine optimal granularity for our anchoring strategy, as shown in Table~\ref{tab:kmeans_effectiveness}. Results show $k=300$ achieving peak performance (MAPE: 22.53\%, $R^2$: 0.5535), representing the optimal balance between cluster specificity and temporal generalization.

Multi-modal Foundation Analysis. Table~\ref{tab:multimodal_ablation} demonstrates the incremental gains from multi-modal integration. User features provide a strong foundation (MAPE: 26.45\%), while textual content significantly enhances performance (MAPE: 24.23\%, R²: 0.5312), confirming that semantic content is highly predictive. Visual and audio modalities further improve performance (final MAPE: 22.53\%, R²: 0.5535), validating our comprehensive multi-modal approach.

\begin{table}[h]
\centering
\resizebox{0.47\textwidth}{!}{%
\begin{tabular}{|c|c|c|c|c|c|c|}
\hline
\textbf{User} & \textbf{Text} & \textbf{Video} & \textbf{Audio} & \textbf{MAPE (↓)} & \textbf{MAE (↓)} & \textbf{R²(↑)} \\
\hline
$\checkmark$ & & & & 26.45\% & 1.6744 & 0.4505 \\
$\checkmark$ & $\checkmark$ & & & 24.23\% & 1.5141 & 0.5312 \\
$\checkmark$ & $\checkmark$ & $\checkmark$ & & 23.87\% & 1.4991 & 0.5382 \\
\textbf{$\checkmark$} & \textbf{$\checkmark$} & \textbf{$\checkmark$} & \textbf{$\checkmark$} & \textbf{22.53\%} & \textbf{1.4677} & \textbf{0.5535} \\
\hline
\end{tabular}%
}
\caption{Ablation study of feature combinations.}
\label{tab:multimodal_ablation}
\vspace{-0.5cm}
\end{table}

Clustering-Based Anchoring Effectiveness. Table~\ref{tab:clustering_effectiveness} validates our TTA-inspired clustering strategy. Raw multi-modal features achieve MAPE 27.17\%, while adding cluster identity improves performance to 26.45\%, demonstrating group recognition value. Critically, incorporating statistical anchors yields substantial improvement (MAPE: 22.53\%, R²: 0.5535), confirming that "how clusters typically perform" provides more stable predictive signals than individual content characteristics—analogous to TTA's neighborhood consistency principles.

\begin{table}[h]
\centering
\resizebox{0.47\textwidth}{!}{%
\begin{tabular}{|l|c|c|c|}
\hline
\textbf{Configuration} & \textbf{MAPE (↓)} & \textbf{MAE (↓)} & \textbf{R²(↑)} \\
\hline
Only multi-modal features & 27.17\% & 1.6769 & 0.4502 \\
+ Cluster identity labels & 26.45\% & 1.6629 & 0.4649 \\
\textbf{+ Statistical anchors} & \textbf{22.53\%} & \textbf{1.4677} & \textbf{0.5535} \\
\hline
\end{tabular}%
}
\caption{Performance comparison of clustering feature configurations.}
\label{tab:clustering_effectiveness}
\vspace{-0.5cm}
\end{table}

Semantic Anchoring Contribution. Table~\ref{tab:semantic_features} evaluates LLM-generated semantic anchors. Our complete AMCFG framework achieves MAPE 22.31\% and MAE 1.4255, showing consistent improvement over statistical anchoring alone. This demonstrates that high-level semantic concepts provide additional temporal stability beyond numerical patterns, validating our dual-anchoring mechanism inspired by TTA's multi-level robustness strategies.

\begin{table}[h]
\centering
\resizebox{0.48\textwidth}{!}{%
\begin{tabular}{|l|c|c|c|}
\hline
\textbf{Configuration} & \textbf{MAPE (↓)} & \textbf{MAE (↓)} & \textbf{R²(↑)} \\
\hline
Baseline (Statistical anchors only) & 22.53\% & 1.4677 & 0.5535 \\
\textbf{AMCFG (+ Semantic anchors)} & \textbf{22.31\%} & \textbf{1.4255} & \textbf{0.5369} \\
\hline
\end{tabular}%
}
\caption{Performance comparison with LLM-generated semantic features.}
\label{tab:semantic_features}
\vspace{-0.5cm}
\end{table}
\subsection{Structural Analysis: Validating Clustering Quality}

\begin{figure}[h]
    \centering
    \includegraphics[width=0.5\textwidth]{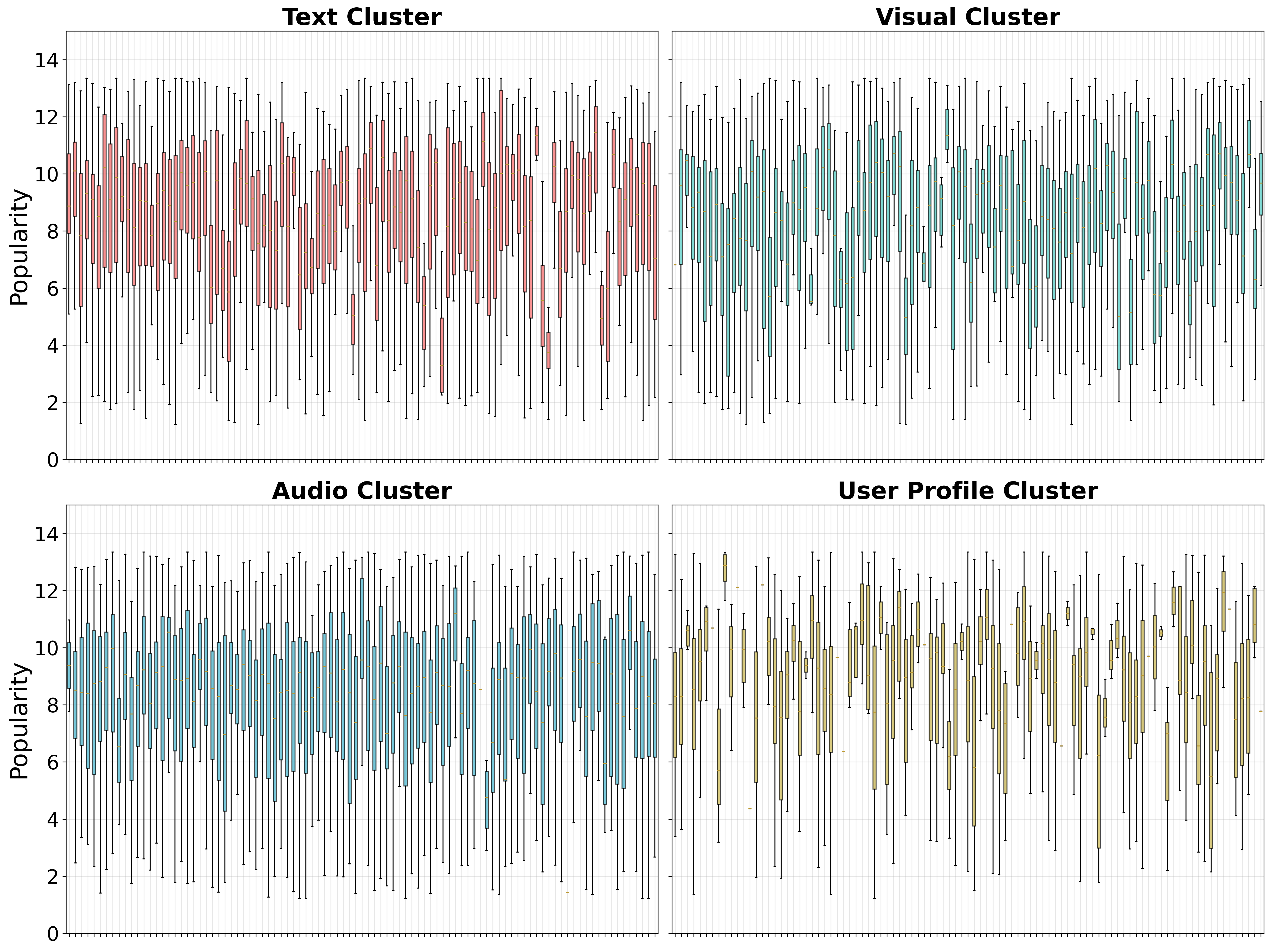}
    \caption{Clustering visualization of four modalities (text, video, audio, and user profile).}
    \label{fig:clustering_viz}
    \vspace{-0.5cm}
\end{figure}

Clustering Visualization. Figure~\ref{fig:clustering_viz} provides visual evidence of our clustering strategy's effectiveness across modalities. The visualization reveals meaningful structures: textual embeddings form semantic neighborhoods, visual features cluster into content styles, audio features group by acoustic patterns, and user profiles segregate into behavioral archetypes. These clear cluster boundaries validate our preprocessing strategies and demonstrate successful multi-modal data partitioning.

\begin{figure}[h]
    \centering
    \includegraphics[width=0.5\textwidth]{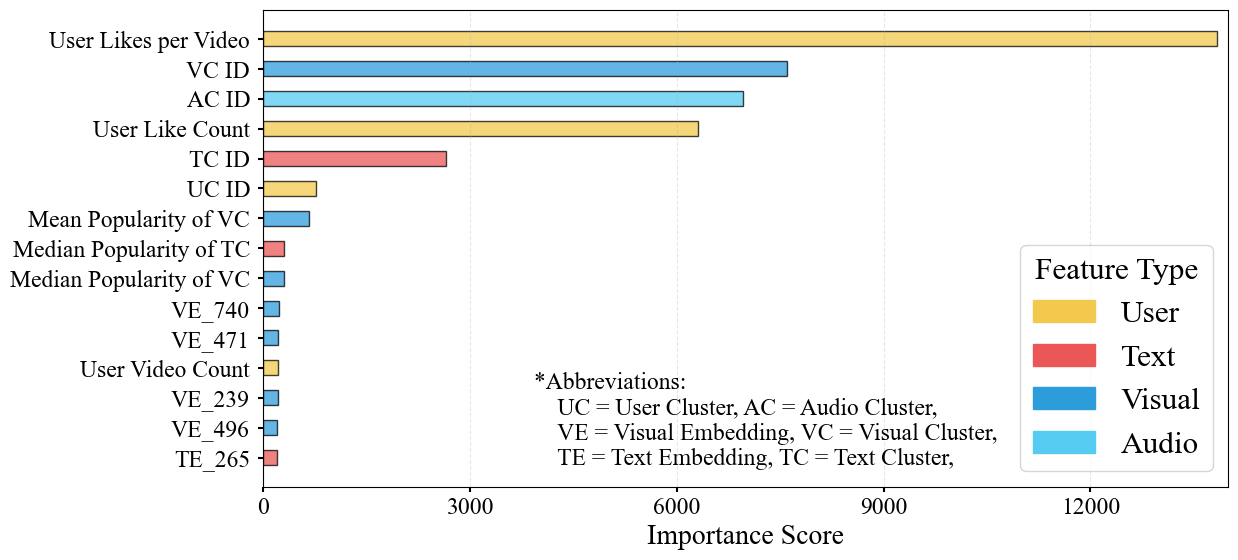}
    \caption{Feature importance analysis across different modalities.}
    \label{fig:feature_importance}
    \vspace{-0.3cm}
\end{figure}

Feature Importance Analysis. Figure~\ref{fig:feature_importance} exposes critical patterns supporting our TTA-inspired design philosophy:

\begin{itemize}
    \item \textbf{Anchor Feature Dominance:} Clustering-derived statistical features consistently rank highest across modalities, confirming that aggregate behavioral patterns provide more temporally stable signals than instance-specific representations.
    
    \item \textbf{User-Centric Patterns:} User metadata features, particularly engagement history, occupy top importance ranks, suggesting creator reputation significantly influences popularity regardless of content specifics.
    
    \item \textbf{Balanced Multi-modal Integration:} Text, video, and audio features distribute throughout importance rankings, indicating successful fusion without single-modality dominance.
\end{itemize}

This empirical evidence strongly supports our framework's core premise: leveraging group-level statistical anchors provides more temporally generalizable features than relying exclusively on individual content embeddings.

\subsection{SMP2025 Challenge Video Track}

To validate practical applicability, we participated in the SMP2025 challenge - Video Track. Table~\ref{tab:smp2025_results} presents our performance on both public and private test sets. The strong performance (MAPE: 17.47\% public, 19.58\% private) with modest generalization gap (2.11\%) confirms that our anchored clustering approach effectively mitigates prediction drift in real-world scenarios, validating the practical utility of our TTA-inspired framework for large-scale deployment.
\begin{table}[h]
\centering
\begin{tabular}{|l|c|c|}
\hline
\textbf{Test Set} & \textbf{MAPE (↓)} & \textbf{Performance Gap} \\
\hline
Public Test & 17.47\% & - \\
Private Test & 19.58\% & +2.11\% \\
\hline
\end{tabular}
\caption{Performance on SMP2025 challenge Video Track.}
\label{tab:smp2025_results}
    \vspace{-0.8cm}
\end{table}
Key Insights. Our experimental results validate three core hypotheses: (1) multi-modal clustering reveals stable behavioral cohorts, (2) statistical anchors provide numerical stability analogous to TTA's neighborhood consistency, and (3) semantic anchors offer conceptual stability complementing statistical patterns. These mechanisms create a robust foundation for temporal generalization in dynamic social media environments.

\section{Conclusion}
Inspired by Test-Time Adaptation's success in addressing distribution shifts through clustering-based anchoring, we introduce our AMCFG framework to combat temporal prediction drift in social media popularity prediction. Our approach establishes a dual-anchoring mechanism discovering temporally-invariant patterns at numerical and conceptual levels: statistical anchors capture stable behavioral patterns within content clusters, while LLM-generated semantic anchors encode high-level thematic concepts transcending superficial trend variations.

Extensive experiments demonstrate this TTA-inspired strategy significantly enhances predictive accuracy and temporal robustness on out-of-distribution data. Our framework bridges rapidly evolving content trends and stable popularity prediction by leveraging structural commonalities rather than surface-level features. This work contributes a novel perspective on temporal generalization, adapting clustering-based feature anchoring from spatial domain adaptation to address temporal distribution shifts in dynamic social media environments.

\begin{table}[h]
  \caption{This prompt table demonstrates the prompt and output examples with VideoLLaMA3 and GPT-4o.}
    \begin{center}
    \rule{\linewidth}{2pt}
    \textbf{\Large VideoLLaMA3}
    \rule{\linewidth}{2pt}
    \end{center}

    \begin{flushleft}
    \textbf{Prompt 1:} "Provide a short description for this video."\\
    \textbf{Output:} "The video showcases a person's hands manipulating various purple objects."\\
    \end{flushleft}

    \begin{flushleft}
    \textbf{Prompt 2:} "Based on the content of this TikTok video, determine its category from the following list: Dance, Comedy, Lip Sync, Tutorial, Beauty \& Fashion, Fitness, Food \& Drink, Pets \& Animals, Vlogging, Challenges, Memes, Technology, Travel, Motivation \& Inspiration, Art \& Creativity, Sports, Music, Social Issues, Unboxing, Pranks, and Others.\\
    \textbf{Output:} "This video is Beauty \& Fashion."\\
    \end{flushleft}

    \begin{flushleft}
    \textbf{Prompt 3:} "Based on the content, style, and context of the video, identify the most likely target audience by describing:

    \begin{itemize}
    \item Age group(s) that would be interested.
    \item Interests or hobbies relevant to the video.
    \item Geographic or cultural background if identifiable.
    \item Reasons or clues from the video that support your audience inference.
    \end{itemize}
    
    Please provide your answer as brief bullet points."
    
    \textbf{Output:} "Age group(s) that would be interested: Men in their 20s to 40s seeking relaxation and self-care. \\(1) - Interests or hobbies relevant to the video: Spa treatments, wellness routines, personal grooming. \\(2) - Geographic or cultural background if identifiable: Likely targeted towards Asian audiences, possibly Chinese, given the use of Chinese characters. \\ (3)- Reasons or clues from the video that support your audience inference: The video showcases a traditional Chinese spa setting with specific treatments like ear cleaning, which is popular in certain Asian cultures. The serene atmosphere and professional services cater to individuals interested in holistic wellness practices."
    \end{flushleft}
    
    \vspace{0.5cm}
    
    \begin{center}
    \rule{\linewidth}{2pt}
    \textbf{\Large GPT-4o}
    \rule{\linewidth}{2pt}
    \end{center}

    \begin{flushleft}
    \textbf{Prompt:} "You are a professional MBTI personality analyst. Based on the following text, internally analyze the author's likely MBTI type by evaluating each of the four dimensions (Extraversion vs. Introversion, Sensing vs. Intuition, Thinking vs. Feeling, Judging vs. Perceiving).
    
    For your internal reasoning, consider the following:
    
    \begin{itemize}
    \item E vs I: Focus on energy orientation (social engagement vs. introspection), tone (outward or inward focus), and mention of interactions with others.
    \item S vs N: Look for language about concrete facts vs. abstract ideas, detail orientation vs. pattern-seeking or imaginative content.
    \item T vs F: Note decision-making language - whether it emphasizes logic/criteria (T) or values/emotions (F).
    \item J vs P: Pay attention to structure, organization, planning (J) vs. spontaneity, flexibility, or openness (P).
    \end{itemize}
    
    Based on your assessment of all four axes, for each group output only the most likely MBTI type (e.g., INFP, ESTJ), and save as .csv (p.s. 0-299, totally 300 group, please give a MBTI for each group).
    \textbf{Output:} This cluster is belong to "INFP"
    
    \end{flushleft}
\end{table}

\newpage
\bibliographystyle{ACM-Reference-Format}
\balance
\bibliography{main} 

\end{document}